\documentclass[modern]{aastex62}

\usepackage{acro}
\usepackage{amsmath}
\usepackage{color}

\newcommand{\dd}{\mathrm{d}}
\newcommand{\diff}[2]{\frac{\dd #1}{\dd #2}}

\newcommand{\fsm}{f_{\mathrm{smooth}}}

\newcommand{\mchirp}{\mathcal{M}}

\newcommand{\MScale}{M_{\mathrm{scale}}}

\newcommand{\wDE}{w_\mathrm{DE}}

\newcommand{\MPISN}{45 \, \MSun{}}
\newcommand{\MTaperScale}{5 \, \MSun{}}

\newcommand{\MScaleOneYear}{43.0^{+1.3}_{-1.3} \, \MSun{}}
\newcommand{\MScaleFiveYear}{44.64^{+0.76}_{-0.81} \, \MSun{}}
\newcommand{\SigmaHPvtOneYear}{6.1 \%}
\newcommand{\SigmaHPvtFiveYear}{2.9 \%}
\newcommand{\SigmaHNaughtTransFiveYear}{2.0}
\newcommand{\zpivot}{0.8}
\newcommand{\wDEOneYear}{-0.68^{+0.17}_{-0.21}}
\newcommand{\wDEFiveYear}{-1.04^{+0.12}_{-0.13}}
\newcommand{\SigmawDEOneYear}{19 \%}
\newcommand{\SigmawDEFiveYear}{12 \%}

\newcommand{\kmsMpc}{\mathrm{km} \, \mathrm{s}^{-1} \, \mathrm{Mpc}^{-1}}
\newcommand{\MSun}{M_\odot}
\newcommand{\perGpcyr}{\mathrm{Gpc}^{-3} \, \mathrm{yr}^{-1}}

\DeclareAcronym{BH}{
  short = BH,
  long = {black hole}
}
\DeclareAcronym{BBH}{
  short = BBH,
  long = {binary black hole}
}
\DeclareAcronym{BNS}{
  short = BNS,
  long = {binary neutron star}
}
\DeclareAcronym{CMB}{
  short = CMB,
  long = {cosmic microwave background}
}
\DeclareAcronym{GW}{
  short = GW,
  long = {gravitational wave}
}
\DeclareAcronym{GWTC1}{
  short = GWTC1,
  long = {Gravitational Wave Transient Catalog 1}
}
\DeclareAcronym{PISN}{
  short = PISN,
  long = {pair instability supernova}
}
\DeclareAcronym{PPISN}{
  short = PPISN,
  long = {pulsational pair instability supernova}
}
\DeclareAcronym{SNR}{
  short = SNR,
  long = {signal to noise ratio}
}

\begin{document}

\title{A Future Percent-Level Measurement of the Hubble Expansion at Redshift
0.8 With Advanced LIGO}

\author[0000-0003-1540-8562]{Will M. Farr}
\affiliation{Department of Physics and Astronomy, Stony Brook University, Stony Brook NY 11794, USA}
\affiliation{Center for Computational Astronomy, Flatiron Institute, 162 5th Ave., New York NY 10010, USA}
\email{will.farr@stonybrook.edu}

\author[0000-0002-1980-5293]{Maya Fishbach}
\affiliation{Department of Astronomy and Astrophysics, University of Chicago, Chicago IL 60637, USA}
\email{mfishbach@uchicago.edu}

\author{Jiani Ye}
\affiliation{Department of Physics and Astronomy, Stony Brook University, Stony Brook NY 11794, USA}
\email{jiani.ye@stonybrook.edu}

\author[0000-0002-0175-5064]{Daniel E. Holz}
\affiliation{Enrico Fermi Institute, Department of Physics, Department of Astronomy and Astrophysics,\\and Kavli Institute for Cosmological Physics, University of Chicago, Chicago IL 60637, USA}
\email{holz@uchicago.edu}

\begin{abstract}
  Simultaneous measurements of distance and redshift can be used to constrain
  the expansion history of the universe and associated cosmological parameters.
  Merging \ac{BBH} systems are standard sirens---their gravitational waveform
  provides direct information about the luminosity distance to the source. There
  is, however, a perfect degeneracy between the source masses and redshift; some
  non-gravitational information is necessary to break the degeneracy and
  determine the redshift of the source.  Here we suggest that the \ac{PISN}
  process, thought to be the source of the observed upper-limit on the \ac{BH}
  mass in merging \ac{BBH} systems at $\sim \MPISN{}$, imprints a mass scale in
  the population of \ac{BBH} mergers and permits a measurement of the
  redshift-luminosity-distance relation with these sources.  We simulate five
  years of \ac{BBH} detections in the Advanced LIGO and Virgo detectors with a
  realistic \ac{BBH} merger rate, mass distribution with smooth \ac{PISN}
  cutoff, and measurement uncertainty. We show that after one year of operation
  at design sensitivity the \ac{BBH} population can constrain $H(z)$ to
  $\SigmaHPvtOneYear$ at a pivot redshift $z \simeq \zpivot$.  After five years
  the constraint improves to $\SigmaHPvtFiveYear$. If the \ac{PISN} cutoff is
  sharp, the uncertainty is smaller by about a factor of two.  This measurement
  relies only on general relativity and the presence of a mass scale that is
  approximately fixed or calibrated across cosmic time; it is independent of any
  distance ladder. Observations by future ``third-generation'' \ac{GW}
  detectors, which can see \ac{BBH} mergers throughout the universe, would
  permit sub-percent cosmographical measurements to $z \gtrsim 4$ within one
  month of observation.
\end{abstract}

\section*{ }

The \ac{GWTC1} contains ten binary black hole merger events observed during
Advanced LIGO and Advanced VIRGO's first and second observing runs
\citep{GWTC-1}. Modeling of this population suggests a precipitous drop in the
merger rate for primary black hole masses larger than $\sim \MPISN{}$
\citep{Fishbach2017,GWTC-1}.  A possible explanation for this drop is the
\ac{PISN} process
\citep{Fowler1964,Rakavy1967,Bond1984,Heger2002,Belczynski2016,Woosley2017,Spera2017,Leung2019}.
This process occurs in the cores of massive stars~\citep[helium core masses $30$--$133
\, \MSun$;][]{Woosley2017} when the core temperature becomes sufficiently
high to permit the production of electron-positron pairs; pair production
softens the equation of state of the core, leading to a collapse which is halted
by nuclear burning \citep{Heger2002}.  The energy produced can either unbind the
star, leaving no \ac{BH} remnant, or drive a mass-loss pulse that reduces the
mass of the star until the \ac{PISN} is halted, leading to remnant masses $\sim
\MPISN{}$ (this latter process is called the \ac{PPISN}). The characteristic
mass of remnant black holes depends weakly on the metallicity of the progenitor
stars; modeling suggests that the upper limit on the remnant mass may vary by
less than 1--2 $\MSun$ for redshifts $0 \leq z \lesssim 2$
\citep{Belczynski2016,Mapelli2017}.  Here we make the conservative choice to
model the effect as a smooth taper in the mass distribution that takes effect
around $m \simeq \MPISN{}$ but acts over a characteristic scale of $\simeq
\MTaperScale$ (see \S\ \ref{sec:simulated-population} for a full description of
our model).  If, in fact, the cutoff is sharper than we assume then our
constraints on cosmology become tighter; a perfectly sharp cutoff reduces our
uncertainty by about a factor of two.

Compact object mergers that emit gravitational waves have a universal
characteristic peak luminosity $c^5/G \simeq 3.6 \times 10^{59} \, \mathrm{erg}
\, \mathrm{s}^{-1}$ that enables direct measurements of the luminosity distance
to these sources, allowing them to be used as ``standard sirens''~\citep{Schutz1986,Holz2005}.
However, the effects of the source-frame mass and redshift are
degenerate in the gravitational waveform; the observed waveform depends only on
the redshifted mass in the detector frame, $m_\mathrm{det} = m_\mathrm{source}
(1 + z)$. General relativity predicts the gravitational waveforms of
stellar-mass \ac{BBH} mergers.  Using parameterized models of these waveforms
\citep{Taracchini2014,Kahn2016,Bohe2017,Chatziioannou2017}, it will be possible
to measure the detector-frame masses with $\sim 20\%$ uncertainty and luminosity
distances \citep{Hogg1999} with $\sim 50\%$ uncertainty for a source near the
detection threshold in Advanced LIGO and Advanced Virgo at design sensitivity
\citep{Vitale2017}.  The relative uncertainty in these parameters scales
inversely with the signal-to-noise ratio of a source.

If we assume that the \ac{BBH} merger rate follows the star formation rate
\citep{Fishbach2018,O1O2Population}, the primary mass distribution follows a
declining power law $m_1^{-\alpha}$ with $\alpha \simeq 0.75$ for $m_1 \lesssim
\MPISN{}$ and tapering off above this mass scale, the mass ratio distribution is
flat, and the three-detector duty cycle is $\sim 50\%$, then Advanced LIGO and
Advanced Virgo should detect $\sim 1000$ \ac{BBH} mergers per year at design
sensitivity over a range of redshifts $0 \leq z \lesssim 1.5$
\citep{O1O2Population}.  The typical detected merger will be at redshift $z
\sim 0.5$.

With this mass distribution, about one in four mergers will have a mass estimate
whose uncertainty provides information about the \ac{PISN} mass scale.  The mass
measurement uncertainty for these events translates directly to an uncertainty
in redshift.  The joint distance-redshift measurement is dominated by the $\sim
50\%$ distance uncertainty for the typical event near the detection threshold,
so the relative uncertainty in the measurement of the expansion rate $H(z)$ at
$z \simeq \zpivot{}$ will be approximately $50 \% / \sqrt{1000/4} \simeq 3 \%$
after one year, and $1.4 \%$ after five years of \ac{BBH} merger observations at
design sensitivity.

\begin{figure}
  \plottwo{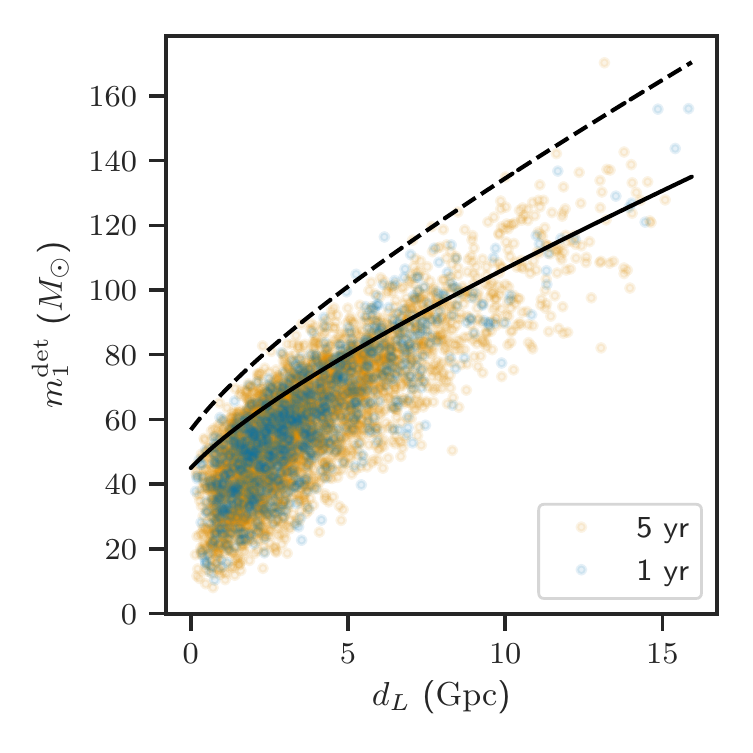}{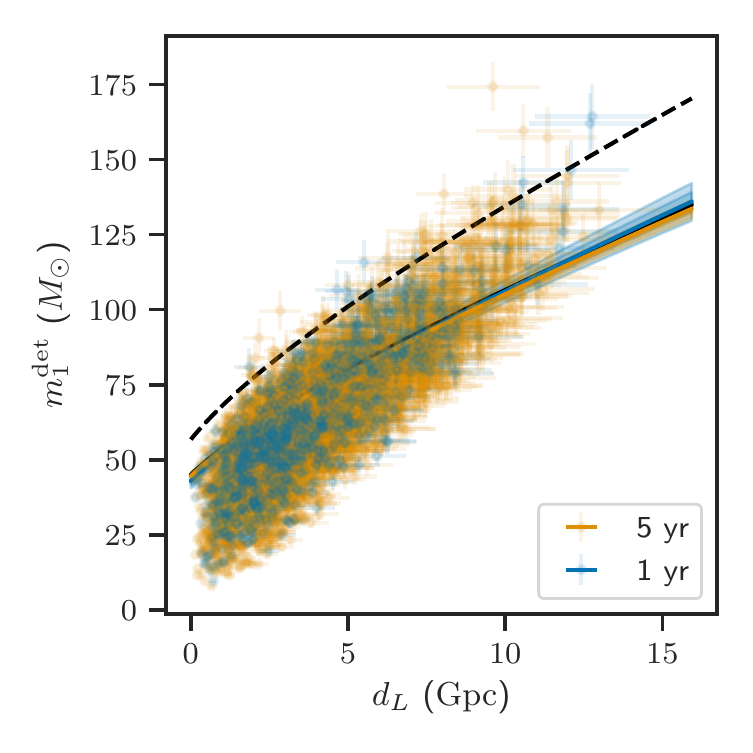}
  \caption{\label{fig:m1-dL} Masses and luminosity distances for a simulated
  population of \ac{BBH} mergers detected by an Advanced LIGO/Virgo network.
  Blue circles denote one year of observations, orange circles five years of
  observations. The solid black line shows the redshifting of the PISN
  detector-frame \ac{BH} mass scale corresponding to the cosmology used to
  generate the events \citep[TT, TE, EE + lowP + lensing + ext]{Planck2016}; the
  dashed black line shows the redshifting of the mass scale at which the
  \ac{PISN} taper has fallen to 1\%.  (Left) The true detector-frame primary
  \ac{BH} masses and luminosity distances.   (Right) The inferred detector-frame
  primary \ac{BH} masses and luminosity distances using our model of the
  measurement uncertainty for each event. Dots denote the mean and bars the
  1$\sigma$ width of the likelihood for each event.  There is a bias in the
  recovery of the masses and distance that becomes more acute at large distances
  due to a failure to model the population (which is not flat in $m_1$ and
  $d_L$) and selection effects in these single-event analyses; our hierarchical
  model that fits the population accounts for these biases. The most-distant
  event biases upward in both mass and distance by several sigma because it
  represents a single ``lucky'' noise fluctuation into detectability out of
  $\sim 2 \times 10^5$ merger events per year within the detector horizon.  We
  also show the inferred distance-mass relation from our analysis of the one
  year and five year mock data sets in the same colors (the solid line gives the
  posterior median, dark band gives the 68\% credible interval, and the light
  band the 95\% credible interval).}
\end{figure}

Detailed calculations are within a factor of two of this back-of-the-envelope
estimate.  We have simulated five years of \ac{GW} observations with Advanced
LIGO and Advanced Virgo at design sensitivity.  We use a local merger rate, mass
distribution, and rate evolution with redshift that are consistent with current
observations \citep{Fishbach2017,Fishbach2018,O1O2Population}.  Our mass
distribution tapers off at $m = \MPISN{}$ to model the effects of the \ac{PISN}
process \citep{Belczynski2016}.  We use a realistic model of the detectability
of sources from this population \citep{GW150914Rate,GW150914RateSupplement} and
for mass and distance estimation uncertainties \citep{Vitale2017}.  The
properties of the simulated population are described more fully in \S
\ref{sec:simulated-population}.  Figure \ref{fig:m1-dL} shows the simulated
detections and uncertainty on detector-frame mass and distance estimates for one
and five years of observation.

\begin{figure}
  \plottwo{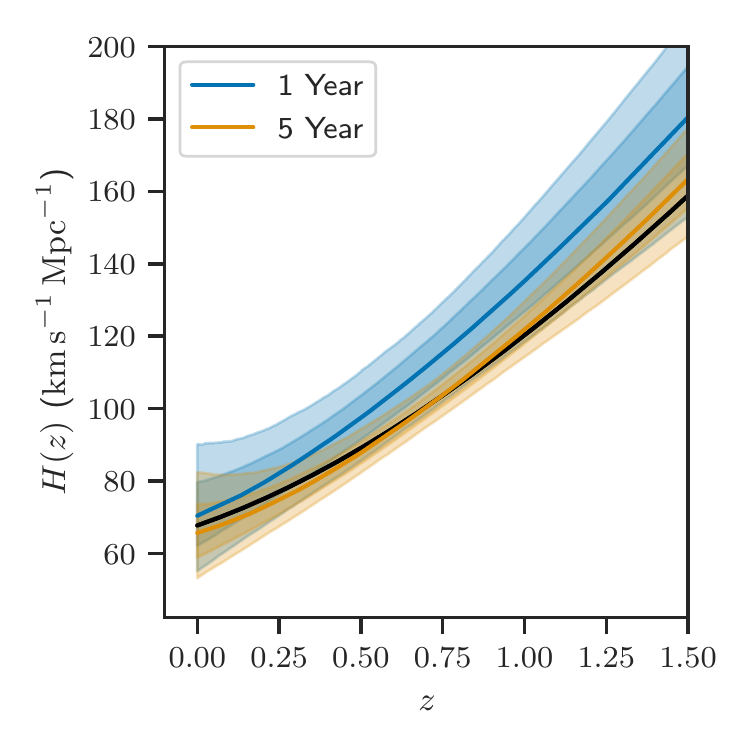}{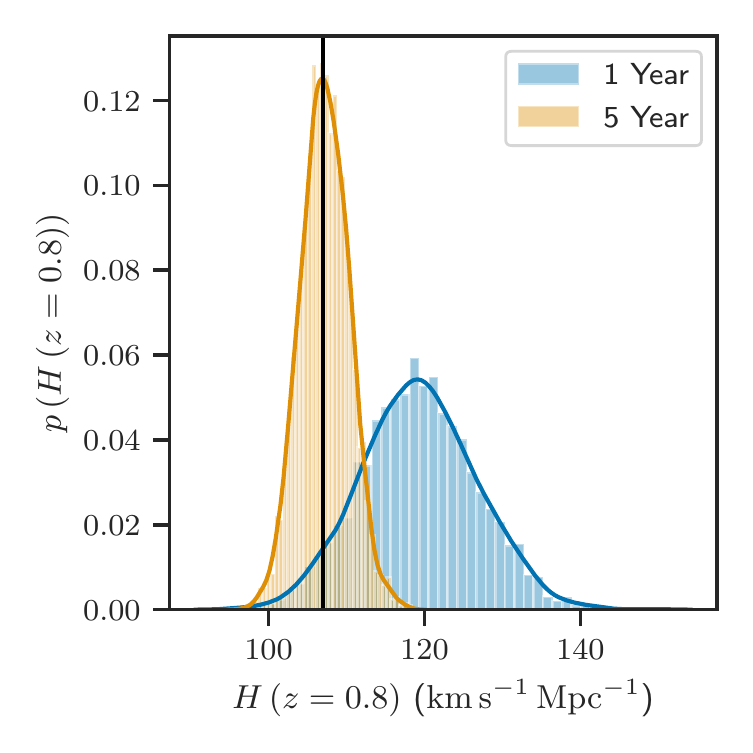}
  \caption{\label{fig:Hz} Inferred cosmological expansion history and distance
  scale.  (Left) The local expansion rate, $H(z)$, inferred from an analysis of
  the one year (blue) and five year (orange) simulated populations using a mass
  distribution model with a parameterized cutoff mass (see text).  The black
  line gives the cosmology used to generate the simulated population \citep[TT,
  TE, EE + lowP + lensing + ext]{Planck2016}.  The solid lines give the
  posterior median $H(z)$ at each redshift; the bands give 1$\sigma$ (68\%) and
  2$\sigma$ (95\%) credible intervals.  The 1$\sigma$ fractional uncertainty on
  $H(z)$ is minimized at $z \simeq \zpivot{}$ for both data sets; after one year
  it is \SigmaHPvtOneYear{} and after five years it is \SigmaHPvtFiveYear{}.
  (Right) Posterior distributions over $H\left(z = \zpivot{}\right)$,
  corresponding to the redshift where the fractional uncertainty is minimized.
  The true $H\left( z = \zpivot{} \right)$ is shown by the black vertical line.
  The posterior after one year is blue, after five years is orange. }
\end{figure}

We fit a parameterized model of the true mass distribution to this data set
accounting for measurement error and selection effects in a hierarchical
analysis \citep{Hogg2010,Mandel2010,Loredo2004,Mandel2019,Farr2019}.  We include
parameters for the power-law slopes in the mass distribution and redshift
evolution, a mass scale and range of masses over which the mass distribution
cuts off due to the \ac{PISN}, a parameterized spatially flat FLRW cosmology
with $H_0$, $\Omega_M$, and $w$ free parameters \citep{Hogg1999}, and parameters
for the true masses and redshift of each detected signal.  The
``population-level'' distribution and cosmological parameters are given broad
priors that are much wider than the corresponding posteriors.  Marginalizing
over all parameters except $H_0$, $\Omega_M$, and $w$ induces a posterior over
expansion histories, $H(z)$, that is shown in Figure \ref{fig:Hz}.  The redshift
at which the fractional uncertainty in $H(z)$ is minimized---the ``pivot''
redshift---is $\zpivot{}$.  After one year of observations, the fractional
uncertainty in $H(z = \zpivot)$ is $\SigmaHPvtOneYear{}$; after five years it is
$\SigmaHPvtFiveYear{}$.  This demonstrates an absolute distance measure to $z
\simeq \zpivot{}$ at percent-level precision; combining this inference on $H(z)$
with other data sets such as observations of baryon acoustic oscillations
\citep{BOSS2015} or Type Ia supernovae \citep{Scolnic2018} can translate this
absolute distance measure to other redshifts (at $z = 0$ it would correspond to
an uncertainty on $H_0$ of $\pm \SigmaHNaughtTransFiveYear{} \, \kmsMpc$)
\citep{BOSS2015,Cuesta2015,Feeney2019}. For example, one can independently
calibrate the Type Ia supernova distance scale without a distance ladder
\citep{Feeney2019,Scolnic2018}, or compare the \ac{GW}-determined distance scale
with one derived from the photon-baryon sound horizon
\citep{Cuesta2015,Aylor2019} in the early universe \citep{Planck2016} or at late
times \citep{BOSS2015}.

Our anticipated constraint on $H(z)$ at the pivot redshift $z \simeq \zpivot$ is
about a factor of two wider than the per-bin anticipated constraint from the
contemporaneous DESI \citep{DESI2016} at comparable redshifts (the combined
constraint from DESI is about a factor of three better than the per-bin
constraint, or a factor of six better than our anticipated measurement).
Comparison of the two measurements would thus enable a percent-level calibration
of the photon-baryon sound horizon scale directly at $z \simeq \zpivot$.

\begin{figure}
  \plotone{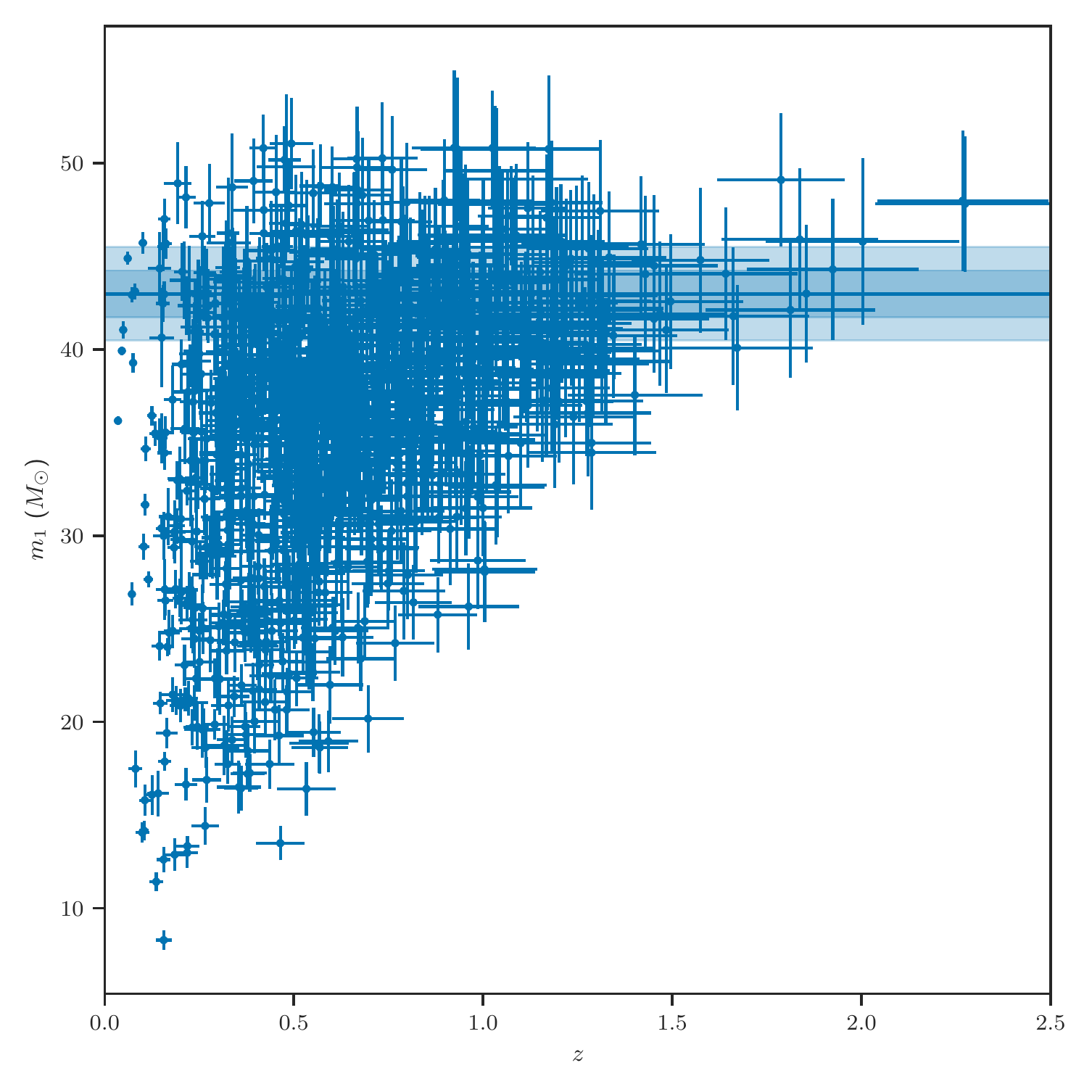}
  \caption{\label{fig:mass-correction} Inferred masses and
  redshifts, and maximum BH mass, for one year of observation.  The points show the posterior mean and 1$\sigma$ (68\%)
  credible ranges for the source-frame primary \ac{BH} masses and redshifts after
  one year of \ac{BBH} merger observations.  The horizontal line is the
  posterior median of the maximum black hole mass set by the PISN process; the
  dark and light bands correspond to the 1$\sigma$ and 2$\sigma$ (68\% and 95\%)
  credible intervals on the maximum mass.  (Compare to Figure \ref{fig:m1-dL}.)
  Our model adjusts cosmological parameters, and therefore the correspondence
  between the measured detector-frame masses and luminosity distances and
  inferred redshifts and source-frame masses, until it achieves a consistent
  upper limit on the source-frame \ac{BH} mass across all redshifts.  After one
  year of synthetic observations we measure $\MScale{} = \MScaleOneYear{}$ (median
  and 68\% credible interval).  After five years (not shown) we measure $\MScale{}
  = \MScaleFiveYear{}$.}
\end{figure}

Our hierarchical model also estimates the source-frame masses and redshifts for
each individual event that incorporate our information about the population.
These results for the one-year data set are shown in Figure
\ref{fig:mass-correction}.  Events pile up near the \ac{PISN} mass scale; in
effect, the cosmology is adjusted so that the measured distances to each event
generate redshifts that produce a constant \ac{PISN} mass scale in the
source-frame from measured detector-frame masses.

\begin{figure}
  \plotone{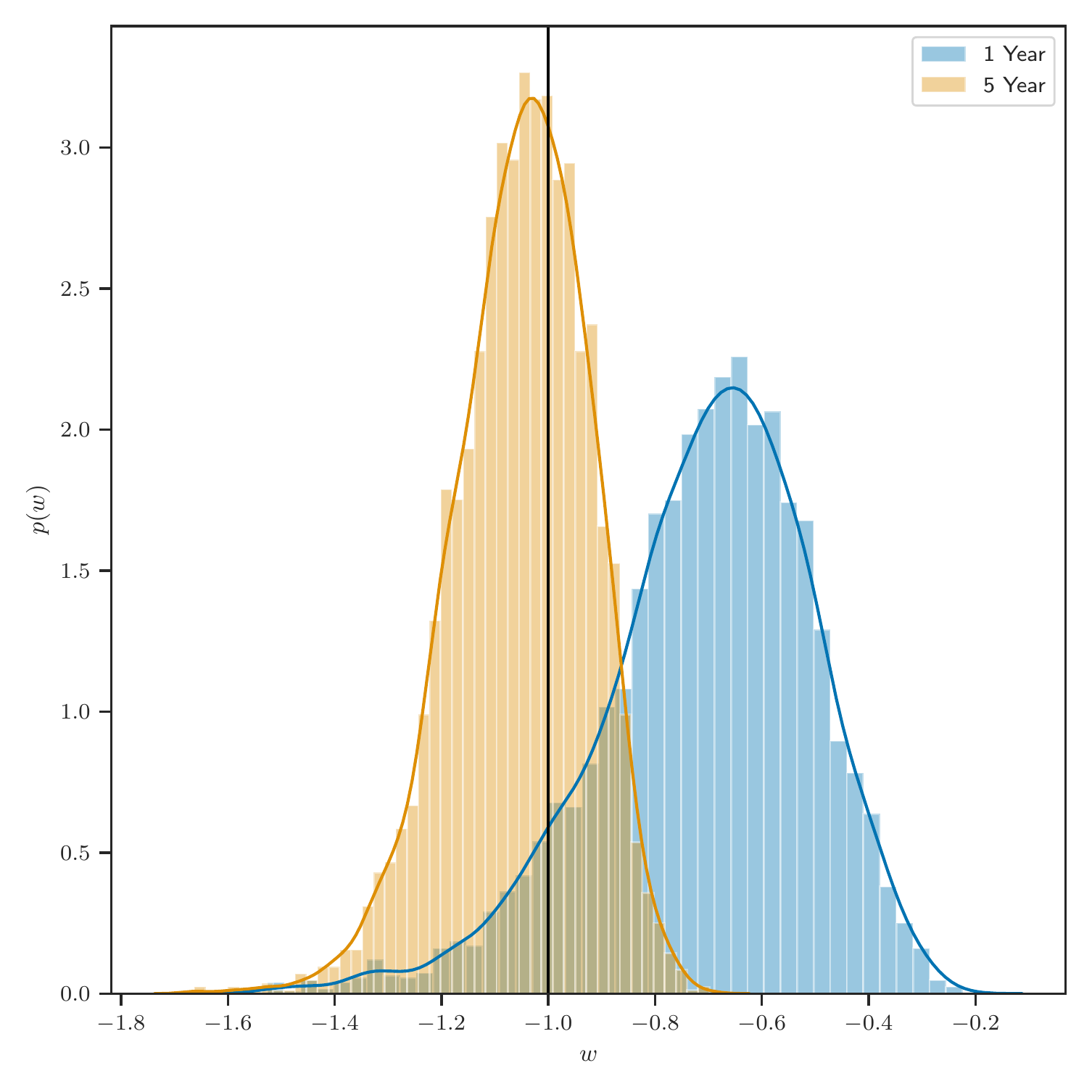}
  \caption{\label{fig:wDE} Posterior on the dark energy equation of state
  parameter after imposing additional cosmological constraints.  If we impose a
  1\% measurement of $H_0$ \citep{Chen2017,Mortlock2018,2018PhRvD..98h3523D} and the constraints on
  $\Omega_M h^2$ from existing observations of the cosmic microwave background
  \citep{Planck2016}, we can infer the equation of state parameter $\wDE{}
  \equiv P_\mathrm{DE} / \rho_\mathrm{DE}$ for dark energy in a $w$CDM
  cosmological model.  (We do not obtain any meaningful constraint on the
  evolution of $\wDE{}$ with redshift when this parameter is allowed to vary, so
  we fix it to a constant across all redshifts.)  We use $\wDE{} = -1$ to
  generate our data set; this value is indicated by the black line above.  The
  posterior obtained on $\wDE{}$ after one year of synthetic observations is
  shown in blue and after five years in orange.  We find $\wDE{} =
  \wDEOneYear{}$ after one year (median and 68\% credible interval) and $\wDE{}
  = \wDEFiveYear{}$ after five years.}
\end{figure}

The pivot redshift for this measurement is close to the redshift where the
physical matter and dark energy densities are equal, and thus this measurement
can be informative about the dark energy equation of state.  If we assume an
independent 1\% measurement of $H_0$~\citep[as could be obtained from \ac{GW}
observations of \ac{BNS} mergers with identified electromagnetic counterparts;][]{Chen2017} and a measurement of the physical matter density at high
redshift~\citep[as obtained by the Planck satellite's measurements of the \ac{CMB};][]{Planck2016}, then the remaining un-constrained parameter in our
cosmological model is $w$, the dark energy equation of state.  Imposing these
additional measurements as a tight prior on the relevant parameters, we find
that our synthetic population of \ac{BBH} mergers can constrain $w$ to
$\SigmawDEOneYear{}$ and $\SigmawDEFiveYear{}$ after one and five years of
observations.  These measurements would be competitive with, but independent
from, other constraints on $w$ \citep[e.g., see][]{2019PhRvL.122q1301A}.  Posteriors for $w$ with these
informative priors are shown in Figure \ref{fig:wDE}.

Our simplistic analysis here assumes that the mass distribution of merging
\acp{BBH} does not change with redshift.  In reality the mass distribution will
change because the metallicity of \ac{BBH} progenitor systems changes with
redshift \citep{Belczynski2016,Mapelli2017}.  The \ac{PISN} mass scale, however,
is not expected to evolve by more than $1$--$2\,\MSun{}$ to $z \simeq 1.5$
\citep{Belczynski2016,Mapelli2017}.  We infer a mass scale in our simple model
of $\MScaleFiveYear{}$ after five years; changes in the \ac{PISN} mass scale for
merging \ac{BBH} systems at a comparable level are a systematic that must be
calibrated to ensure an accurate measurement.  \ac{BBH} mergers thus become
``standardizable sirens.''

There is a possibility that the \ac{PPISN} process, a sequence of incomplete
pair-instability-driven mass loss events, could lead to a pile-up of \ac{BBH}
systems near the upper mass limit
\citep{Belczynski2016,Marchant2018,Talbot2018}.  Current LIGO observations are
inconclusive about the existence of such a ``pile up'' in the mass distribution
\citep{O1O2Population}.  Should one exist, it would offer another mass scale in
the mass distribution that could improve upon the constraints presented here. It
may also be possible to detect and calibrate evolution in the \ac{PISN} limit by
comparing the location and amplitude of the pile up as a function of distance,
since these properties would respond differently to a change in the \ac{PISN}
mass scale.

The possible existence of so-called ``second generation'' \ac{BBH} mergers~\citep[mergers where one black hole is itself a merger product, see e.g.][]{2017ApJ...840L..24F} could fill in the
\ac{PISN} mass gap, but are not expected to be prevalent enough to obscure the
falloff in the mass distribution due to the \ac{PISN} limit discussed here
\citep{Rodriguez2019}.

It is likely that by the mid 2020s there will be two gravitational wave
detectors operating in addition to the two LIGO and one Virgo detectors
\citep{ObsScenarios}; additional detectors do not dramatically improve distance
or mass estimates \citep{Vitale2017}, but the higher SNR afforded from the
additional detectors could extend the detection horizon leading to a factor of
$\sim 4$ increase in the number of \ac{BBH} detections and a resulting factor of
two improvement in the constraints presented here.  Third generation \ac{GW}
detectors, planned for construction in the mid-2030s, would detect $\sim 15,000$
\ac{BBH} mergers per month to $z \gtrsim 10$, with a typical relative
uncertainty on $d_L$ of $\sim 10 \%$ at $z \simeq 2$ \citep{Vitale2018}.
Provided the \ac{PISN} mass scale is properly calibrated, such detectors could
achieve sub-percent uncertainty in cosmography to high redshifts $z \lesssim 5$
with \emph{one month} of \ac{BBH} merger observations.

\acknowledgments

We thank Stephen Feeney for providing a sounding board for the methods discussed
in this paper.  We thank Jon Gair for providing a thorough and helpful LIGO
Scientific Collaboration internal review.  We thank Daniel Mortlock and Jakub
Scholtz for suggestions that improved Figure 1.  We thank Risa Wechsler for
discussions of DESI at the Aspen Center for Physics.  We acknowledge the 2018
April APS Meeting and Barley's Brewing Company in Columbus, OH, USA where this
work was originally conceived. MF was supported by the NSF Graduate Research
Fellowship Program under grant DGE-1746045. MF and DEH were supported by  NSF
grant PHY-1708081. They were also supported by the Kavli Institute for
Cosmological Physics at the University of Chicago through NSF grant PHY-1125897
and an endowment from the Kavli Foundation. DEH also gratefully acknowledges
support from the Marion and Stuart Rice Award.  WMF thanks the Aspen Center for
Physics where this work was completed; it is supported by National Science
Foundation grant PHY-1607611.

All code and data used in this analysis, including the \LaTeX{} source for this
document, can be found under an open-source license at
\url{https://github.com/farr/PISNLineCosmography}.

\software{Numpy \citep{numpy},
Scipy \citep{scipy},
IPython \citep{IPython},
Matplotlib \citep{Matplotlib},
scikit-learn \citep{SciKitLearn},
astropy \citep{astropy:2013,astropy:2018},
PyStan \citep{Stan,PyStan},
Seaborn \citep{Seaborn},
Arviz \citep{Arviz} 
}

\bibliography{pisn-line}

\appendix

\section{Simulated Population}
\label{sec:simulated-population}

We draw our synthetic observations from a population that follows
\begin{multline}
  \label{eq:population}
  \diff{N}{m_1 \dd m_2 \dd V \dd t} = \frac{R_{30}}{\left( 30 \, \MSun \right)^2} \left( \frac{m_1}{30 \, \MSun} \right)^{-\alpha} \left( \frac{m_2}{30 \, \MSun} \right)^{\beta} \left( 1 + z \right)^{\gamma} \\ \times \fsm\left( m_1 \mid m_l, \sigma_l, m_h, \sigma_h \right) \fsm\left( m_2 \mid m_l, \sigma_l, m_h, \sigma_h \right),
\end{multline}
where all quantities are evaluated in the comoving frame and
\begin{equation}
  \label{eq:smooth}
  \fsm\left( m \mid m_l, \sigma_l, m_h, \sigma_h \right) = \Phi\left( \frac{\log m - \log m_l}{\sigma_l} \right) \left[ 1 - \Phi\left( \frac{\log m - \log m_h}{\sigma_h} \right) \right]
\end{equation}
is a function that tapers smoothly to zero when $m \lesssim m_l$ or $m \gtrsim
m_h$ over a scale in log-mass of $\sigma_l$ and $\sigma_h$; $\Phi(x)$ is the
standard normal cumulative distribution function.  (We enforce $m_2 \leq m_1$.)

\begin{figure}
  \plotone{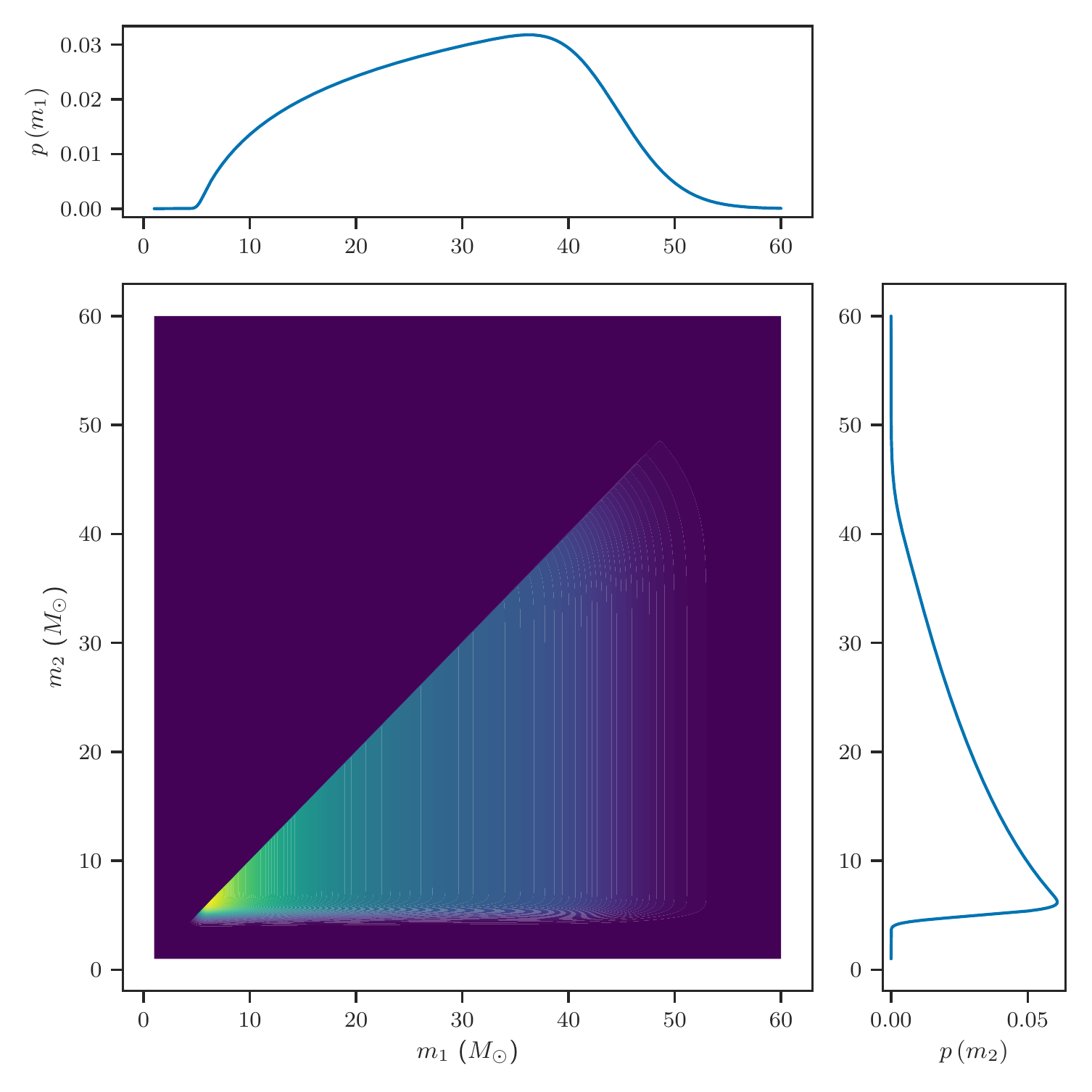}
  \caption{\label{fig:marginal-masses} \textbf{Mass distributions.} The joint
  and  marginal mass distributions for the masses in merging \ac{BBH} systems
  implied by the merger rate density in Eq.\ \eqref{eq:population} and the
  parameter choices in Eq.\ \eqref{eq:parameters}. The turnover at $m \simeq
  \MPISN{}$ due to the \ac{PISN} mass scale is apparent in the primary mass
  distribution.}
\end{figure}

We have chosen population parameters that are consistent with the current
observations reported in GWTC-1 \citep{GWTC-1,O1O2Population}:
\begin{eqnarray}
  \label{eq:parameters}
  R_{30} & = & 64.4 \\
  \alpha & = & 0.75 \\
  \beta & = & 0.0 \\
  \gamma & = & 3.0 \\
  m_l & = & 5 \, \MSun \\
  m_h & = & 45 \, \MSun \\
  \sigma_l & = & 0.1 \\
  \sigma_h & = & 0.1.
\end{eqnarray}
with these choices the volumetric merger rate at $z = 0$ is $60 \, \perGpcyr$.
The corresponding marginal mass distributions for $m_1$ and $m_2$ are shown in
Figure \ref{fig:marginal-masses}.

We model the cosmology as a flat wCDM model which depends on the Hubble
constant, $H_0$, the matter density in units of the critical density,
$\Omega_M$, and the dark energy equation of state parameter $w$
\citep{Hogg1999}.

\section{Measurement and Selection Model}
\label{sec:measurement}

After drawing a catalog of true merger parameters from the population
distribution in \S\ \ref{sec:simulated-population}, we use an approximation
\citep{Fishbach2018} to the true measurement and selection process in a
gravitational wave detector \citep{Veitch2015}.

We assume a 50\% duty cycle for our detector network.

We use IMRPhenomPv2 waveforms \citep{Hannam2014} to compute the expected optimal
\ac{SNR}, $\bar{\rho}\left( m_1, m_2, d_L \right)$, in an Advanced LIGO detector
operating at design sensitivity \citep{ObsScenarios} for the sources in our
catalog assuming that the source appeared in a face-on configuration directly
above the detector.  We draw a random number $\Theta \in [0,1]$ from a
distribution that results from averaging \ac{GW} signal amplitude over position
on the sky and binary orientations \citep{Finn1993}.  The observed \ac{SNR} in a
single Advanced LIGO detector follows
\begin{equation}
  \rho \sim N\left( \bar{\rho} \Theta, 1 \right).
\end{equation}
We approximate the detectability of a source in a three-detector network as a
threshold on the observed single-detector \ac{SNR}, only including sources in
our detected catalog if $\rho > 8$ \citep{GW150914RateSupplement}.

For detected sources we assume the symmetric mass ratio,
\begin{equation}
  \eta \equiv \frac{m_1 m_2}{\left( m_1 + m_2 \right)^2},
\end{equation}
and chirp mass in the detector frame,
\begin{equation}
  \mchirp \equiv \left( m_1 + m_2 \right) \eta^{3/5} \left( 1 + z \right),
\end{equation}
are measured with uncertainty
\begin{equation}
  \eta_\mathrm{obs} \sim N\left( \eta, 5 \times 10^{-3} \frac{8}{\rho} \right)
\end{equation}
and
\begin{equation}
  \log \mchirp_\mathrm{obs} \sim N\left( \log \mchirp, 3 \times 10^{-2} \frac{8}{\rho} \right),
\end{equation}
where the observed symmetric mass ratio is constrained to $0 \leq
\eta_\mathrm{obs} \leq 0.25$.  The angular amplitude factor, $\Theta$, is
measured with uncertainty
\begin{equation}
  \Theta_\mathrm{obs} \sim N\left( \Theta, 5 \times 10^{-2} \frac{8}{\rho} \right)
\end{equation}
constrained to $0 \leq \Theta_\mathrm{obs} \leq 1$.  Our complete observed data
for each detection in the catalog consists of $\rho$, $\mchirp_\mathrm{obs}$,
$\eta_\mathrm{obs}$, and $\Theta_\mathrm{obs}$; the likelihood function for
$m_1$, $m_2$, $\Theta$, and $d_L$ given these data follows from the above
distributions and uncertainties (which are assumed to be measured for each
event).

The uncertainty on the angular amplitude factor, $\Theta$, is tuned to reproduce
the correct distribution of detected distance uncertainties for a three-detector
network at design sensitivity \citep{Vitale2017}. Our model reproduces the
correlated mass measurements, scaling with \ac{SNR}, and typical uncertainty in
mass and distance space that would result from a fuller analysis of detected
systems at much lower computational cost than full parameter estimation
\citep{Veitch2015}.

\section{Hierarchical Analysis}
\label{sec:hierarchical}

Our hierarchical analysis proceeds from a model of a censored Poisson process
with measurement uncertainty \citep{Loredo2004,Mandel2019}.  The joint posterior
on the parameters of each source, $\theta_i \equiv \left\{ m_1^{(i)}, m_2^{(i)},
\Theta^{(i)}, d_L^{(i)} \right\}$, and the population-level parameters, $\lambda
= \left\{ R_{30}, \alpha, \beta, \gamma, m_l, m_h, \sigma_l, \sigma_h, H_0,
\Omega_M, w \right\}$ (see \S\ \ref{sec:simulated-population}) given
observational data $d_i$ ($i = 1, \ldots, N_\mathrm{obs}$) is
\begin{equation}
  \label{eq:hierarchical-model}
  \pi \left( \lambda, \left\{ \theta_i \right\} \mid \left\{ d_i \right\} \right) = \prod_{i=1}^{N_\mathrm{obs}} \left[ p\left( d_i \mid \theta_i \right) \diff{N}{\theta_i}(\lambda) \right] \exp\left[ - \Lambda(\lambda) \right] p\left( \lambda \right),
\end{equation}
where $p\left( d \mid \theta \right)$ is the likelihood function representing
the measurement process detailed in \S\ \ref{sec:measurement};
$\diff{N}{\theta}(\lambda)$ is the population model described in \S\
\ref{sec:simulated-population}; $\Lambda$ is the expected number of detections
given population parameters $\lambda$,
\begin{equation}
  \label{eq:selection-integral}
  \Lambda(\lambda) \equiv \int_{\rho > 8} \dd d \, \dd \theta \, p\left( d | \theta \right) \diff{N}{\theta}\left(\lambda \right);
\end{equation}
and $p(\lambda)$ is a prior on the population parameters.  For our default
analysis, we choose prior distributions for each population parameter that are
much wider than the corresponding posterior; when constraining $w$, the dark
energy equation of state, we impose a tight prior on $H_0$ and $\Omega_M$ as
described in the main text.

Though $p\left( d \mid \theta \right)$ is computable in closed form for our
simplified measurement model, we implement this function as a Gaussian mixture
model density estimate over samples $\theta$ drawn from $\theta \sim p\left( d
\mid \theta \right)$ \citep{SciKitLearn}.  Thus our implementation of the
hierarchical model is agnostic to the form of the likelihood function, and can
easily consume samples from a full parameter estimation analysis over real
\ac{GW} observational data \citep{GWTC-1}.

Similarly, though $\Lambda(\lambda)$ is computable in closed form for our
simplified selection model, we estimate the integral in Eq.\
\eqref{eq:selection-integral} via importance sampling from a sample of
``detected'' systems as described in \S\ \ref{sec:measurement} drawn from a
reference population \citep{Farr2019}.  Thus, our analysis could deal with a
selection function from a real search over \ac{GW} data, represented as a list
of synthetic signals drawn from a reference population that have been
successfully detected by a search pipeline.

We use the Hamiltonian Monte-Carlo sampler Stan \citep{Stan} to sample from the
distribution over the high-dimensional parameter space of the $\theta_i$ and
$\lambda$ defined in Eq.\ \eqref{eq:hierarchical-model}.  Our samplings involve
four independent chains of 1000 samples, pass convergence tests based on the
$\hat{R}$ statistic \citep{Gelman1992}, and we have verified that each parameter
has an effective sample size that is at least $100$ (and greater than $1000$ for
most parameters).

\end{document}